\documentclass[aps,prd,showpacs,twocolumn]{revtex4}

\usepackage{graphicx}

\newcommand{\anp}{g_{np}}
\newcommand{\anS}{g_{n\Sigma}}
\newcommand{\apL}{g_{p\Lambda}}
                                                                                
\newcommand{\manp}{\left(1-\anp^2\right)}
\newcommand{\panp}{\left(1+\anp^2\right)}
\newcommand{\mapL}{\left(1-\apL^2\right)}
\newcommand{\papL}{\left(1+\apL^2\right)}
\newcommand{\manS}{\left(1-\anS^2\right)}
\newcommand{\panS}{\left(1+\anS^2\right)}
\newcommand{\en}{\mu_n}
\newcommand{\ep}{\mu_p}
\newcommand{\eL}{\mu_\Lambda}
\newcommand{\pL}{k_\Lambda}
\newcommand{\eS}{\mu_\Sigma}
\newcommand{\pS}{k_\Sigma}

\begin{document}

\title{$R$-modes of accreting hyperon stars as persistent sources of
gravitational waves}
\author{Mohit Nayyar}
\author{Benjamin J. Owen}
\affiliation{Institute for Gravitational Physics and Geometry and Center for
Gravitational Wave Physics, Department of Physics, The Pennsylvania State
University, University Park, PA 16802-6300}
\date{$$Id: paper.tex,v 1.92 2006/03/22 04:46:31 nayyar Exp $$}

\begin{abstract}

The $r$-modes of accreting neutron stars could be a detectable source of
persistent gravitational waves if the bulk viscosity of the stellar matter
can prevent a thermal runaway.
This is possible if exotic particles such as hyperons are present in the
core of the star.
We compute bulk viscous damping rates and critical frequencies for
$r$-modes of neutron stars containing hyperons in the framework of
relativistic mean field theory.
We combine the results of several previous calculations of the
microphysics, include for the first time the effect of rotation, and
explore the effects of various parameters on the viability of persistent
gravitational wave emission.
We find that persistent emission is quite robust, although it is 
disfavored
in stars below 1.3--1.5~$M_\odot$ depending on the equation of state.
In some cases persistent emission is compatible with temperatures as low as
$10^7$~K, observed in some accreting neutron stars in quiescence.
\end{abstract}

\preprint{IGPG-05/12-1}

\maketitle

\section{Introduction}

The $r$-modes (fluid oscillations governed by the Coriolis force) of
rapidly rotating neutron stars have attracted much interest as possible
sources of gravitational waves and mechanisms for regulating the spins of
neutron stars.
See Ref.~\cite{Stergioulas:2003yp} for a recent review of the many physical
and astrophysical issues related to the $r$-modes; here we focus on
gravitational wave emission.
Gravitational radiation drives the $r$-modes unstable and could lead to
detectable gravitational wave emission in two scenarios.
In one scenario, a newborn neutron star could radiate a substantial
fraction of its rotational energy and angular momentum as gravitational
waves changing in frequency on a timescale of a year or
more~\cite{Owen:1998xg}.
In the other scenario, $r$-modes in rapidly accreting neutron stars in
low-mass x-ray binaries (LMXBs) could be persistent sources of periodic
gravitational waves~\cite{Bildsten:1998ey, Andersson:1998qs}.
(Here the $r$-modes provide a specific mechanism for a more general torque
balance argument~\cite{Papaloizou:1978, Wagoner:1984pv}).

Currently the latter scenario for gravitational wave emission (from LMXBs)
looks like a brighter prospect for detection.
There is now evidence from several approaches that the amplitude of an
$r$-mode growing due to the instability is limited by nonlinear fluid
dynamics to a relatively small value~\cite{Schenk:2001zm, Morsink:2002ut,
Arras:2002dw, Brink:2004bg, Brink:2004qf, Brink:2004kt, Sa:2004gn}.
While low-amplitude long-lived $r$-modes in newborn neutron stars still can
lead to astrophysically interesting effects such as the regulation of
spins, a low mode amplitude renders the gravitational wave signal
undetectable unless there is a very nearby supernova~\cite{Owen:1998xg}.
Also, if neutron stars contain particles more exotic than neutrons and
protons---such as hyperons, where an up or down quark in a nucleon is
replaced by a strange quark---there are additional viscous damping
mechanisms which may eliminate the instability altogether in very young,
hot neutron stars~\cite{Langer:1969, Jones:1970, Jones:1971, Jones:2001ie,
Jones:2001ya, Lindblom:2001hd, Haensel:2001em, vanDalen:2003uy} or strange
quark stars~\cite{Madsen:1998qb}.
However, for the LMXB scenario, low amplitude is not a
problem~\cite{Bildsten:1998ey, Andersson:1998qs} and the additional
viscosity actually renders stars with exotic particles better candidates
for gravitational wave detection.
The reason for the latter is subtle, and bears explanation.

\begin{figure}
\centerline{\includegraphics[width=2.8in]{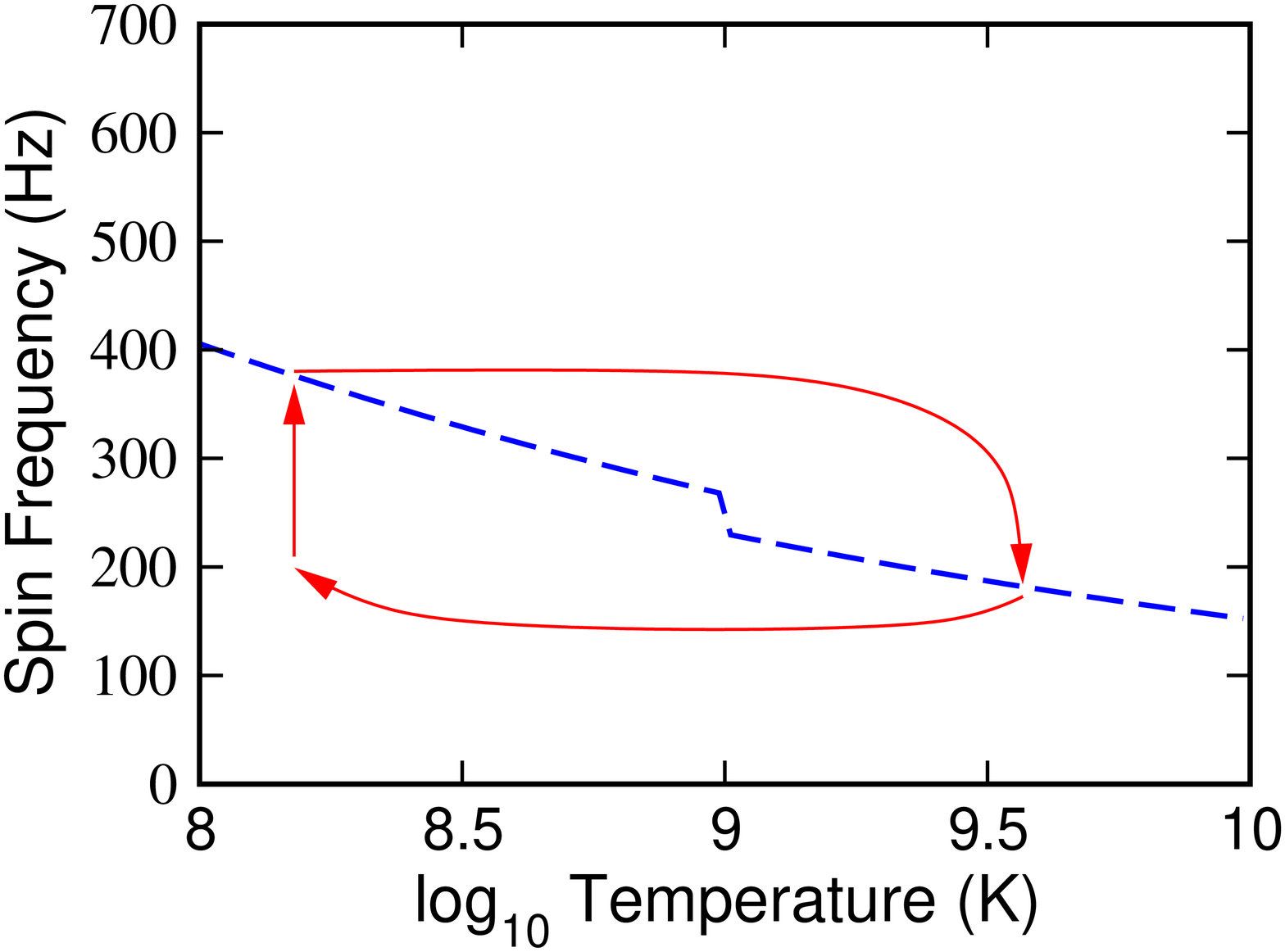}}
\centerline{\includegraphics[width=2.8in]{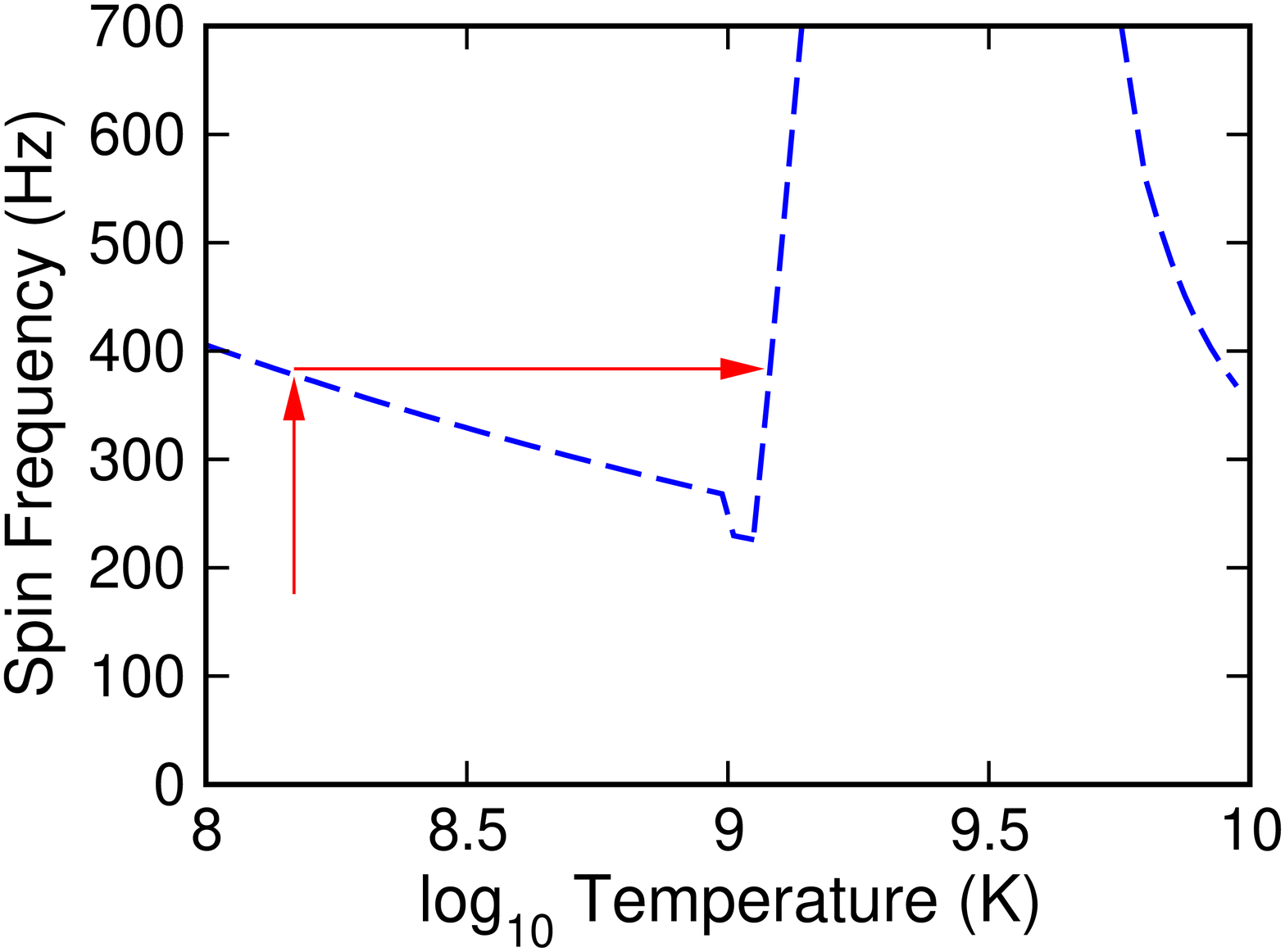}}
\caption{
\label{runaway}
Critical frequency curves are given (qualitatively) by the dashed lines.
The top plot includes no bulk viscosity due to hyperons or other strange
particles.
In this case an accreting neutron star traversing the loop indicated
undergoes a thermal runaway and has a low gravitational radiation duty
cycle.
The bottom plot includes hyperon bulk viscosity.
In this case the thermal runaway is blocked, and an accreting star is a
source of persistent gravitational waves as it remains in equilibrium at
the last arrowhead.
}
\end{figure}

Real neutron stars are not perfect fluids, and thus viscous (and other)
damping mechanisms compete with gravitational wave driving of the
$r$-modes.
The strengths of the driving and damping mechanisms can be expressed as
timescales which depend on the rotation frequency and temperature of the
star (usually assumed to be nearly isothermal).
Therefore it is useful to consider the location of a star in the
temperature-frequency plane, as shown in Fig.~\ref{runaway}.
The strength of viscosity can be graphically represented by a curve in that
plane that is the locus of all points where the driving and damping
timescales are equal---this defines a critical frequency as a function of
temperature.
Since the driving timescale decreases with frequency, stars above the
critical frequency curve have one or more unstable $r$-modes, while stars
below it are stable and stars on it are marginally stable.

Figure~\ref{runaway} plots examples of critical frequency curves in the
temperature range $10^8$~K$ \ll T \ll 10^{10}$~K appropriate for the LMXB
emission scenario.
(The needed range is higher than most observed temperatures of LMXBs in
quiescence because of a thermal runaway; see below.)
At low temperatures the damping in Fig.~\ref{runaway} is taken to be
dominated by shear viscosity in a boundary layer between the solid crust
and fluid core (from Ref.~\cite{Lindblom:2000gu}, augmented by a constant
relative crust-core velocity in the range discussed in
Ref.~\cite{Levin:2000vq}).
There are many other possible curves for the low temperature part of the
plot, corresponding to more complicated damping mechanisms such as
turbulence~\cite{Wu:2000qy} or superfluid magnetoviscous
effects~\cite{Kinney:2002mq}, but they generally share the qualitative
property of decreasing with temperature.
(Some low-temperature curves lie above the observed range of spins
entirely, but the observed spins of low-mass x-ray binaries are easier to
explain if the curve lies in the region indicated~\cite{Bildsten:1998ey,
Andersson:1998qs, Chakrabarty:2003kt}.)
At high temperatures the damping is probably dominated by bulk viscosity,
either from the Urca process (perturbation of $\beta$-equilibrium) or from
nonleptonic processes involving strange particles such as hyperons.
The Urca process, which requires no exotic particles, does not affect the
critical frequency curve for $T < 10^{10}$~K and thus does not show up in
Fig.~\ref{runaway}.
Thus the top plot in Fig.~\ref{runaway} shows a critical frequency curve
for a neutron star, and the bottom plot shows such a curve for a star with
hyperons (with the high-temperature part of the curve derived from
Ref.~\cite{Lindblom:2001hd}).
This type of plot looks encouraging for gravitational wave detection
because there is plenty of room above the curve for stars to be unstable
and thus emitting gravitational waves.

However, the gravitational wave emission duty cycle could be much smaller
than 100$\%$ due to a thermal runaway~\cite{Levin:1998wa, Spruit:1998mk}.
This happens generically when the critical frequency decreases with
temperature.
In that case, plotted at the top of Fig.~\ref{runaway}, a star will
execute a loop as shown and radiate only during the time it spends above
the curve.
A stable star which begins at the bottom left of the loop is spun up by
accretion until it moves above the critical frequency and the instability
is triggered.
The shear from the growing $r$-modes then causes the star to heat up,
moving it rightwards on the loop.
As the temperature rises, so does the rate of neutrino cooling, causing the
star to drop down toward the critical frequency again at a temperature of a
few times $10^9$~K.
After falling below the critical frequency, the $r$-mode heating is removed
and the star drifts leftward along the bottom of the loop until returning
to its initial position.
Although the time it takes for a star to complete the loop is dominated by 
the accretion rate, the timescale for gravitational wave emission depends 
mainly on the saturation amplitude of $r$-mode oscillations.
For a saturation amplitude ($\alpha$ in the notation of
Ref.~\cite{Lindblom:1998wf}) of order unity, the duty cycle for
gravitational wave emission is of the order $10^{-6}$~\cite{Levin:1998wa}.
Whereas the duty cycle can be as high as about $30\%$ for the lowest
predicted values of the saturation amplitude ($\alpha\simeq10^{-5}$), for
typical estimates of the saturation amplitude ($\alpha\simeq10^{-4}$) the
duty cycle is only of order $10^{-1}$~\cite{Heyl:2002pe}.
Advanced LIGO will be able to detect at most one LMXB (Sco X-1) without
narrowbanding (and hurting its ability to see other sources), or 6--7 LMXBs
with narrowbanding around a series of different
frequencies~\cite{Cutler:2002me}.
The small number of detectable systems and the fact that the timescale for
a star to complete a loop is much longer than a human lifetime mean that a
duty cycle of order $10^{-1}$ or less is pessimistic for gravitational wave
searches for the $r$-modes.

If the critical frequency increases with temperature as in the bottom plot
of Fig.~\ref{runaway}, the thermal runaway can be blocked.
A rapidly accreting star in an LMXB can have a duty cycle of order unity
for emission of gravitational radiation as it sits on the curve or makes
small peregrinations about it~\cite{Wagoner:2001kx}.
(It could also keep emitting, although less detectably, for some time after
the rapid accretion shuts off~\cite{Reisenegger:2003cq}.)
This could happen for stars which exhibit a rise in the critical frequency
curve at a low enough temperature.
The rise is typical for stars where high bulk viscosity processes
involving hyperons~\cite{Wagoner:2002vr, Wagoner:2003vi} or strange
quarks~\cite{Andersson:2001ev} are at work.
Thus the case for the gravitational wave emission scenario is in fact
strengthened by high bulk viscosity from processes which are fundamentally
quark-quark interactions.
One might expect similar critical frequency curves to arise for stars
containing other forms of strange matter such as a kaon condensate or
mixed quark-baryon phase, but this has yet to be investigated.
The thermal runaway is blocked if the increase in temperature (width of
the loop in Fig.~\ref{runaway}) is enough to take the star from the
negatively sloped part of the instability curve to the positively sloped
part.
The larger the saturation amplitude of $r$-modes the greater is the
increase in the temperature.
Whether the star makes the jump to the positively sloped curve also
depends on which cooling mechanisms are operative, and on the shape of the
low-temperature instability curve, which in turn depends on which damping
mechanism dominates at low temperatures.
Notwithstanding the wide range of estimates for $\alpha$, the shape of the
negatively sloped curve, and the cooling mechanisms, the question of the
$r$-modes in LMXBs as a persistent source of gravitational radiation comes
down to asking whether there is a rise in the instability curve around
$10^9$~K or lower.

Our purpose in this paper is to revisit the question of whether there is a
rise in the critical frequency curve around $10^9$~K or lower, as in the
bottom of Fig.~\ref{runaway}, leading to persistent gravitational wave
emission.
We focus (for now) on neutron stars containing hyperons (hyperon stars),
because hyperons are in some sense the most conservative and robust of the
many proposals for exotic matter in the cores of neutron stars:
Some properties of hyperons can be measured in the laboratory, both in
vacuum and in the environment of a light nucleus, which combined with
astronomical observations allows one to constrain some of the many
uncertainties in building an equation of state~\cite{Glendenning:1997wn,
Lackey:2005tk}.
Equations of state which allow for hyperons generally produce them at
densities relevant for neutron stars, about twice nuclear density and up.

We synthesize and extend results of previous work on this topic.
In arguing the case for persistent gravitational wave emission,
Wagoner~\cite{Wagoner:2002vr, Wagoner:2003vi} and Reisenegger and
Bonacic~\cite{Reisenegger:2003cq} base their critical frequency curves on
bulk viscosity coefficients obtained by combining the results of Lindblom
and Owen~\cite{Lindblom:2001hd} (hereafter LO) and Haensel, Levenfish, and
Yakovlev~\cite{Haensel:2001em} (hereafter HLY).
The LO and HLY viscosities were obtained by different calculations and
produced somewhat different results.
LO used a detailed self-consistent model of a multi-component fluid
described by relativistic mean field theory~\cite{Glendenning:1997wn}, but
since they were primarily concerned with the high-temperature regime
appropriate to newborn neutron stars they treated the effect of
superfluidity with a rough approximation.
LO also made some errors which resulted in a bulk viscosity coefficient a
factor of 20 or more too high.
HLY were more careful with superfluidity, using consistent damping factors
in the collision integrals, but instead of evaluating reaction rates and
thermodynamic derivatives within relativistic mean field theory they used
``order of magnitude estimates'' of some quantities which resulted in more
than an order of magnitude disagreement with the microscopic results of LO
(and with Jones~\cite{Jones:2001ya}).
We correct some mistakes in LO and combine their self-consistent
microphysical model with a more careful treatment of superfluidity similar
to that of HLY.
We also treat the macroscopic physics more carefully than LO, including the
effect of rotation on stellar structure which in some cases can
significantly affect damping timescales and critical frequency curves.
(HLY made only order of magnitude estimates of mode damping timescales and
did not plot critical frequency curves.)

We also address the question ``Does the viability of persistent
gravitational wave emission require fine tuning of parameters?''
Many of the parameters that go into building the equation of state have
significant uncertainties, and those that go into computing reaction rates
and bulk viscosities are even more uncertain.
We investigate the viability of persistent gravitational wave emission with
respect to variation of several microphysical numbers such as hyperon
coupling constants and the superfluid bandgap, and even which nonleptonic
reaction is most important (taking into account the results of van Dalen
and Dieperink~\cite{vanDalen:2003uy}).
We also investigate the dependence on the mass of the star, since cooling
observations~\cite{Yakovlev:2004iq} and timing of radio pulsars in
binaries~\cite{Nice:2005fi, Ransom:2005ae} indicate a wider mass range than
the traditionally assumed clustering around 1.4~$M_\odot$.
(Cooling observations also might be interpreted to favor the existence of
strange particles such as hyperons, although the data still can be fit by
exotic cooling from purely nucleonic matter.)

The organization of the rest of the paper is as follows. 
In Sec.~II we describe the microphysical model which leads to the equation
of state and ultimately the macroscopic coefficient of bulk viscosity.
In Sec.~III we plot the critical frequency curves for a range of neutron
star masses and microphysical parameters.
In Sec.~IV we summarize our findings and discuss possible improvements.

\section{Microphysics}

We model the composition of a neutron star within the framework of
relativistic mean field theory industrialized by Glendenning and described
in Ref.~\cite{Glendenning:1997wn}.
Here we do not consider meson condensates or exotica such as a mixed
quark-baryon phase, but do allow for the presence of hyperons which are
somewhat constrained by laboratory and astronomical data.

\subsection{Equation of state}

There are several free parameters in this framework, some of which are
better known than others.
We use a range of parameters set by combining the old laboratory
constraints with recent astronomical observations~\cite{Lackey:2005tk}.
At nuclear density (2--3$\times10^{15}$~g/cm$^3$) the matter is, as in all
models, composed of neutrons, protons, electrons, and muons.
As the density rises hyperons generically appear, their order of appearance
(and density thresholds) changing somewhat with the precise set of
parameter values but roughly corresponding to $\Sigma^-$, $\Lambda$,
$\Sigma^0$, $\Sigma^+$, $\Xi^-$, starting at roughly twice nuclear density.
Other baryons typically appear in this framework only at densities higher
than those found in the cores of the most massive neutron stars, nearly 10
times nuclear density.
At such high densities the asymptotic freedom of the strong nuclear force
is likely to lead to quark matter anyway, and any baryonic model must be
considered somewhat suspect.
As we shall see, most of the astrophysically important dissipation comes
from lower densities (2--3 times nuclear density) where this deficiency of
the model is not important, i.e. low-energy effective field
theory is good enough.
The framework of relativistic mean field theory has some advantages over
others, such as an inherently causal equation of state even at high
densities, but has some disadvantages including the neglect of correlations
between particles which should become more important at high densities.

In this framework, the strong interaction between baryons is modeled as a
tree-level exchange of isoscalar mesons $\sigma$, vector mesons $\omega$,
and isovector mesons $\rho$.
The effective Lagrangian includes kinetic and tree-level interaction terms
for the baryons, leptons, and mesons, as well as an effective potential for
the $\sigma$ expanded up to $O(\sigma^4)$.
The variables of the theory are the Fermi momenta $k_i$ of the baryons and
leptons, supplemented by the field strengths of the $\sigma$, $\omega$, and
$\rho$ mesons.
Expansion coefficients and coupling constants can be fit to numbers
extracted from measurements of nuclei and hypernuclei at saturation density.
The fitting process (and Lagrangian) is described in detail in
Ref.~\cite{Glendenning:1997wn}.
We use an updated set of fit parameters from Ref.~\cite{Lackey:2005tk}.

For $N$ species of baryons, two of leptons, and three of mesons, the
composition of the matter consists of $N+5$ unknowns (the baryon, lepton
and meson fields) which must be determined by $N+5$ equations.
The first three equations are the Euler-Lagrange equations for the mesons
determined by varying the effective Lagrangian of
Ref.~\cite{Glendenning:1997wn}.
The equations are rendered tractable by assuming that the fields are given
by their mean values in a uniform static ground state, and take the form
\begin{eqnarray}
\label{firstEuler}
{\omega_0} &=& {\sum_B}{\frac{g_{\omega B}}{m_\omega^2}}{n_B},
\\
{\rho_{03}} &=& {\sum_B}{\frac{g_{\rho B}}{m_\rho^2}}I_{3
B}{n_B},
\\                                                                              
{m_\sigma^2}{\sigma} &=& {\sum_B}{g_{\sigma
B}}{\rho_{s,B}}-b m_{N}{g_\sigma}{(g_\sigma
\sigma)}^2-c{g_\sigma}{(g_\sigma \sigma)}^3.
\end{eqnarray}
Here the index $B$ labels baryon species, $\omega_0$ is the timelike
component of the vector meson field (the spatial components vanish), $n_B$
is the number density of baryon $B$, and $\rho_{03}$ is the isospin
3-component of the timelike component of the isovector meson field.
Throughout this Section we use units such that $\hbar=c=1$.
The scalar density is given by
\begin{equation}
{\rho_{s,B}}=\sum_B \frac{2 J_B+1}{2\pi^2}
 \int_0^{k_B}
\frac{m_B^{\star 2}(\sigma)}{\sqrt{k^2+m_B^{\star 2}(\sigma)}}
 k^2 dk.
\end{equation}
The coupling constants between the baryon B and the mesons are given by
$g_{\sigma B}$, $g_{\omega B}$, and $g_{\rho B}$.
We assume, as in Ref.~\cite{Glendenning:1997wn}, that they are given by one
set of values $(g_\sigma, g_\omega, g_\rho)$ for the nucleons and another
$(x_\sigma g_\sigma, x_\omega g_\omega, x_\rho g_\rho)$ for the hyperons,
so that the $x$'s measure relative coupling strengths for hyperons.
The coupling constants for the self-interaction terms of the scalar field
are $b$ and $c$.
The masses of the mesons are given by $m_{\sigma}$, $m_{\omega}$, and
$m_{\rho}$.
The quantities for the $B$th baryon $I_{3B}$ and $J_B$ are, respectively, 
its
3-component of isospin, and its spin.
The Fermi momentum of the baryon species is given by $k_B$ while its 
effective mass is given by $m_B^* = m_B - g_{\sigma B} \sigma$.
The quantity $m_N$ represents the average mass of a nucleon and is used to
make $b$ and $c$ dimensionless.
The constraints of charge neutrality and conservation of baryon number
(if a particular baryon number density $n$ is assumed) provide two more
equations which can be written, respectively, as
\begin{eqnarray}
\label{consQ}
\sum_i n_i q_i &=& 0,
\\
\label{consB}
\sum_B n_B &=& n.
\end{eqnarray}
Here $n_i$ denotes the number density of fermion species $i$ and $q_i$ is
its electric charge in units of $e$.
Now we are down to $N$ unknowns, which can be eliminated with the $N$
equations of generalized $\beta$-equilibrium
\begin{equation}
\label{lastEuler}
\mu_i=b_i\mu_n-q_i\mu_e.
\end{equation}
(The neutrinos are assumed to have zero chemical potential.)
Here $\mu_i$ is the chemical potential of fermion species $i$ and $b_i$
is its baryon charge.
(Actually there are $N+2$ equations of $\beta$-equilibrium including the
leptons, but the equations for the electron and neutron are identities and
thus do not count toward the elimination of unknowns.)
The chemical potential of each baryon species is given by 
\begin{equation}
\mu_B   = g_{\omega B}\omega_0 +g_{\rho B}\rho_{03}I_{3B}
        +\sqrt{k_B^2+m^{*2}_B},
\end{equation}
while that of each lepton species is given by
\begin{equation}
\mu_L   = \sqrt{k_L^2+m^{2}_L}.
\end{equation}
Here $k_L$ and $m_L$ are, respectively, the Fermi momentum and mass of the 
$L$th leptonic species.

For a given baryon number density $n$, the baryon, meson, and lepton fields
are found by simultaneously solving
Eqs.~(\ref{firstEuler})--(\ref{lastEuler}).
The numerical technique is based on a multi-dimensional root finder and
thus requires an initial guess to ensure finding the correct root.
We first solve for a low value $n$ below saturation density where the
system can be approximated as a weakly interacting Fermi gas and analytic
approximations can be found as in Ref.~\cite{Glendenning:1997wn}.
The code then steps up in baryon density $n$, at each step using the
previous step's values for the unknowns as the initial guess for the root
finder.
Once the fields are found for a given $n$, the mass-energy density
$\epsilon$ and pressure $p$ can be determined:
\begin{eqnarray}
{\epsilon} &=& 
{\frac{1}{3}bm_{N}\left(g_{\sigma}\sigma\right)^3} + 
{\frac{1}{4}c\left(g_{\sigma}\sigma\right)^4} + 
{\frac{1}{2}}m_{\sigma}^2\sigma^2
\nonumber\\
&&+{\frac{1}{2}}m_{\omega}^2
\omega_0^2 +{\frac{1}{2}}m_{\rho}^2\rho_{03}^2
\nonumber\\
&& +\sum_{B}{\frac{1}{\pi^2}} \int_0^{k_B} \sqrt{k^2+m_B^{*2}}k^2dk
\nonumber\\
&& +\sum_{L}{\frac{1}{\pi^2}} \int_0^{k_L} \sqrt{k^2+m_L^2}k^2dk,
\\
\label{pressure}
p &=& 
-{\frac{1}{3}bm_{N}\left(g_{\sigma}\sigma\right)^3} -
{\frac{1}{4}c\left(g_{\sigma}\sigma\right)^4}
-{\frac{1}{2}}m_{\sigma}^2\sigma^2
\nonumber\\
&&+\frac{1}{2}m_{\omega}^2\omega_0^2 
+\frac{1}{2}m_{\rho}^2\rho_{03}^2
\nonumber\\
&& +\sum_{B}\frac{1}{\pi^2} \int_0^{k_B} \frac{k^4dk}{\sqrt{k^2+m_B^{*2}}}
\nonumber\\
&& +\sum_{L}\frac{1}{\pi^2} \int_0^{k_L} \frac{k^4dk}{\sqrt{k^2+m_L^2}}.
\end{eqnarray}
Here the summations are over the baryon species (B) and the lepton species
(L).
These expressions can be combined to produce tabulations of $p(\epsilon)$,
the equation of state.

There remains the problem of choosing the values of the constants
$(g_\sigma/m_\sigma)^2$, $(g_\omega/m_\omega)^2$, $(g_\rho/m_\rho)^2$, $b$,
$c$, $x_\sigma$, $x_\omega$, and $x_\rho$.
Each choice of constants produces a different equation of state with
different hyperon threshold densities, neutron star maximum mass, etc.
We use five equations of state from~\cite{Lackey:2005tk} chosen to be
compatible with recent measurements of neutron star
masses~\cite{Nice:2005fi, Ransom:2005ae} and a gravitational
redshift~\cite{Cottam:2002cu} as well as hypernuclear
data~\cite{Glendenning:1997wn}.
The constants in Eqs.~(\ref{firstEuler})--(\ref{lastEuler}) are fit in the
manner of Glendenning~\cite{Glendenning:1997wn} to a $\Lambda$ binding of
-28~MeV, saturation density 0.153~fm$^{-3}$, binding energy per nucleon
-16.3~MeV at saturation, and isospin asymmetry coefficient 32.5~MeV, as
well as a range of incompressibilities $K$, nucleon effective masses $m^*$,
and hyperon couplings $x_\sigma$.
Since it is easiest to qualitatively understand the equations of state in
terms of the latter three parameters, we give them in Table~\ref{eostab}.
Low $K$ and high $m^*$ lead to a soft equation of state, while high $K$ and
low $m^*$ lead to a stiff equation of state.
Since most of the matter in the star is at high density, the hard core
repulsion dominates the strong interaction.
Thus high $x_\sigma$ postpones hyperon formation to higher densities,
reducing hyperon populations (and thus viscosity) and stiffening the
equation of state.

\begin{table}
\caption{\label{eostab}
Parameters for five relativistic mean field equations of state used in this
paper.
}
\begin{ruledtabular}
\begin{tabular}{lccc}
Name & $K$ (MeV) & $m^*/m_N$ & $x_\sigma$
\\
\hline
H3 & 300 & 0.70 & 0.60
\\
H4 & 300 & 0.70 & 0.72
\\
H5 & 300 & 0.80 & 0.66
\\
H6 & 240 & 0.70 & 0.67
\\
H7 & 240 & 0.80 & 0.68
\end{tabular}
\end{ruledtabular}
\end{table}

All of these equations of state are somewhat different from that used by LO
and HLY.
Though both HLY and LO use the ``case 2'' equation of state from an older
paper by Glendenning~\cite{Glendenning:1985jr}, there were misprints in
that paper which were corrected by HLY but not by LO.
The scalar self interaction coupling constants $b$ and $c$ in
Ref.~\cite{Glendenning:1985jr} should be $3$ and $4$ times larger, 
 respectively, than the values that are quoted for them.
Also, case 2 of Ref.~\cite{Glendenning:1985jr} used $K=285$~MeV and
$x_\sigma = x_\omega = x_\rho = \sqrt{2/3}$, and had several minor
differences in other constants (we use values from the 2002 Particle Data
Group review~\cite{Hagiwara:2002fs}).
As a result, the newer equations of state we use have higher hyperon
populations and thus higher viscosities.

\subsection{Relaxation timescale}

The reactions that contribute most to bulk viscosity are the weak
interaction processes, as their relaxation timescales are within a few
orders of magnitude of the $r$-mode period (milliseconds).

Among these, the most significant that can be calculated from first
principles are the non-leptonic weak interactions involving the lightest
hyperons, $\Sigma^-$ and $\Lambda$, as they occur at lower densities than
the more massive particles and thus have higher populations in a given
star.
Following LO, we calculate matrix elements for the reactions
\begin{eqnarray}
\label{Sigm}
n + n &\leftrightarrow& p+{\Sigma^-}
\\
\label{Lamb}
n+p &\leftrightarrow& p+{\Lambda}
\end{eqnarray}
as tree-level Feynman diagrams involving the exchange of a W boson.
The latter reaction was also considered (with a phenomenological free
parameters) by HLY.
These reactions are combined to calculate an overall microscopic relaxation
timescale.
(This is the timescale on which a small perturbation of the neutron
fraction returns to its equilibrium value.)

In addition to these reactions, Jones~\cite{Jones:2001ya} includes
\begin{equation}
n+n\leftrightarrow n+\Lambda
\label{not_included_reac}
\end{equation}
which is the dominant nonleptonic process observed in $\Lambda$ hypernuclei
in the laboratory.
We do not consider this reaction as it has no contribution based on a
W-boson exchange, even though the bare-mass interaction rate for this
process is known.
Using Eq.~(15) from Jones~\cite{Jones:2001ya} we estimate that
reaction~(\ref{not_included_reac}) has a relaxation timescale that is
longer than the timescale of the $\Lambda$ reaction~(\ref{Lamb}) by a
factor of about 2 and thus does not change the final viscosity much.
(When accounting for the macroscopic structure of the star, the $\Sigma^-$
process dominates the overall bulk viscosity since the $\Sigma^-$
population extends to lower densities and therefore a greater volume
fraction of the star than the $\Lambda$ population.)
There are various other concurrently occuring processes which contribute
to the net reaction rate but their rates are not easy to predict.
Even these rates are subject to substantial uncertainties:
Van Dalen and Dieperink~\cite{vanDalen:2003uy} use a meson-exchange model
for all three hyperon reactions and find reaction rates of order 10--100
times greater than LO and thus bulk viscosity coefficients 10--100 times
lower than LO.
(As we shall see in the next section, such a change has less of an effect
on the $r$-mode critical frequency than one might think.)
Our result for the net rate is then a lower limit, and consequently an
upper limit on the bulk viscosity, in the typical range of temperatures
for LMXBs.

To summarize the results of LO, the relaxation timescale can be computed
as a function of the equilibrium matter fields.
Neglecting superfluidity, the result is
\begin{equation}
\label{tau1}
{1\over\tau} = {(kT)^2 \over 192\pi^3} k_\Sigma \langle
|\mathcal{M}_\Sigma^2| \rangle {\delta\mu \over n\, x_n}
\end{equation}
when the $\Lambda$ hyperons have not yet appeared and
\begin{equation}
\label{tau2}
{1\over\tau} = {(kT)^2 \over 192\pi^3} \left( k_\Sigma \langle
|\mathcal{M}_\Sigma^2| \rangle + k_\Lambda \langle
|\mathcal{M}_\Lambda^2| \rangle \right) {\delta\mu \over n\, x_n}
\end{equation}
when both $\Lambda$ and $\Sigma^-$ are present.
Here $kT$ is Boltzmann's constant times the temperature.
The matrix elements $\mathcal{M}$ are computed as tree-level Feynman
diagrams, squared, summed over initial spinors, and averaged over the
angular part of the collision integral (indicated by $\langle\rangle$).
(Note that van Dalen and Dieperink~\cite{vanDalen:2003uy} do not angle
average.)
They are functions of the Fermi momenta and meson fields given by
Eqs.~(4.28) and~(4.29) of LO,
\begin{widetext}
\begin{eqnarray}
\label{msqL}
\langle |{\cal M}_\Lambda|^2 \rangle &=& {G_F^2 \sin^22\theta_C \over 15}
\left\{ 120\manp\mapL m_n^* m_p^{*2} m_\Lambda^*
- 20\manp\papL m_n^* m_p^* \left(3\ep\eL
- \pL^2\right) \right.\nonumber
\\
&&\left. - 10\panp
\mapL m_p^* m_\Lambda^* \left(6\en\ep - 3k_n^2 + \pL^2\right) + 2\left[
\panp\papL + 4\anp\apL \right] \right.\nonumber\\
&&\left.\times\left[ 5\ep\eL \left( 6\en\ep + 3k_n^2 -
\pL^2 \right) + \pL^2
\left( 10\en\ep + 5k_n^2 + 10k_p^2 - \pL^2 \right) \right]
+ \left[ \panp\papL - 4\anp\apL \right]\right. \nonumber\\
&&\left.\times\left[ 10\en\eL \left( 6m_p^{*2} +
3k_n^2 + \pL^2 \right)
+ \pL^2 \left( -20\ep^2 + 15k_p^2 - 3\pL^2 + 5{(k_n^2 -
k_p^2)^2 /( k_p^2 - \pL^2)} \right) \right] \right\},
\\
\label{msqS}
\langle |{\cal M}_\Sigma|^2 \rangle &=& {2\over15} G_F^2 \sin^22\theta_C
\left\{ 180\manp\manS m_n^{*2} m_p^* m_\Sigma^* - 40\manp\panS m_n^* m_p^*
 \left( 3\en\eS
- \pS^2 \right) - 20\right.\nonumber
\\
&& \times\panp\manS m_n^* m_\Sigma^* \left( 6\en\ep - 3k_p^2 + \pS^2 \right)
- 5\manp\manS m_p^* m_\Sigma^* \left( 6\en^2 + 6k_n^2 - 3k_p^2 -\pS^2 \right)
\nonumber
\\
&& +4\left[ \panp\panS + 4\anp\anS \right] \left[ 10\en^2 \left( 3\ep\eS +
\pS^2 \right) + 5\ep\eS \left( 6k_n^2 - 3k_p^2 - \pS^2 \right) + \pS^2
\left( 10k_n^2 + 5k_p^2 \right.\right. \nonumber
\\
&& \left.\left.
- \pS^2 \right) \right]
 + \left[ \panp\panS - 4\anp\anS \right] \left[ -10m_n^{*2} \left(
3\ep\eS + \pS^2 \right) + 10\en \left( 6\en\ep\eS - 2\ep\pS^2 - 3\eS k_p^2
\right.\right.
\nonumber
\\
&& \left.\left.\left. + \eS\pS^2 \right) + 15k_p^2\pS^2 - 3\pS^4
\right] \right\}.
\end{eqnarray}
\end{widetext}
We reproduce them here since the latter equation in LO is missing some
terms (although the code used to generate the LO results is not missing
them).
Also, LO actually used kinetic energies instead of chemical potentials
which should be used to be consistent with the quasiparticle picture of
Fermi liquid theory.
Correcting this reduces the relaxation time by a factor of 10 or more,
making up much of the difference with the results of van Dalen and
Dieperink~\cite{vanDalen:2003uy} who use a drastically different model of
the interaction.
In the matrix elements, $G_F$ is the Fermi constant and $\theta_C$ is the
Cabibbo angle.
The $g$'s are axial-vector coupling constants, which we take to have their
values measured from vacuum $\beta$-decay of particles at
rest~\cite{Hagiwara:2002fs}.
The remaining factor on the far right of Eqs.~(\ref{tau1}) and~(\ref{tau2})
is explained in the next subsection.

HLY, in contrast, used the nonrelativistic limit
\begin{equation}
\langle |{\cal M}_\Sigma|^2 \rangle
= 4G_F^2 \sin^22\theta_C \left(1 + 3\anp\anS\right)^2
\label{msqS_nonrel}
\end{equation}
of Eq.~(\ref{msqS}) and left the term in parentheses as a phenomenological
parameter, which they simply set to 0.1.
This is somewhat justified because the Particle Data group values for the
couplings $\anp$ and $\anS$ result in a near-perfect cancellation of the
term.
Any in-medium change of these values would have a disproportionately large
effect as a result.
We (and LO) find that including the full Eq.~(\ref{msqS}) erases the effect
of the near-cancellation of the leading-order terms and produces a running
value of the term which can increase by more than an order of magnitude
above 0.1.

\subsection{Bulk viscosity}

We use for $\zeta$, the macroscopic coefficient of bulk viscosity, the
special relativistic expression derived by Lindblom and
Owen~\cite{Lindblom:2001hd}:
\begin{equation}
\zeta = {p(\gamma_\infty - \gamma_0)\tau \over 
1+\left(\hat\omega\tau\right)^2}.
\label{zeta}
\end{equation}
where $p$ is the pressure,
$\hat\omega$ is the angular frequency of the $r$-mode in a frame corotating
with the star, and $\tau$ is the net microscopic relaxation time we just
computed.
The fast adiabatic index $\gamma_{\infty}$ (for infinite-frequency
perturbations) can be written as
\begin{equation}
\gamma_\infty = \sum_i
\frac{n_i}{n}\left(\frac{\partial p}{\partial n_i}\right),
\label{gammai}
\end{equation}
where $n_i = k_i^3/(3\pi^2)$ is the number density of species $i$ and the
partial derivatives can be evaluated explicitly from Eq.~(\ref{pressure}).
The slow adiabatic index $\gamma_0$ (for zero-frequency perturbations)
can be written as
\begin{equation}
\gamma_0 = \left({\frac{n}{p}}\right) \left(\frac{dp}{dn}\right)
\label{gamma0},
\end{equation}
and is straightforward to evaluate for example by differentiating the
equation of state.
However, while LO derived the correct relativistic expression and noted
that it reduces the factor $\gamma_\infty - \gamma_0$ by a factor of
2--2.5 from the nonrelativistic expression, their Fig.~2 and code for the
rest of the paper incorrectly used a data file with the nonrelativistic
expression and thus are a factor 2--2.5 too high.

Now we address the factor $\delta\mu/ \delta x_n$ in Eqs.~(\ref{tau1})
and~(\ref{tau2}) in the same manner as LO.
This factor is determined by the constraints of charge and baryon
conservation~(\ref{consQ}) and~(\ref{consB}), plus the constraint that the
reaction
\begin{equation}
\label{strong}
n + \Lambda \leftrightarrow p^+ + \Sigma^-
\end{equation}
is in equilibrium since its reaction rate is many orders of magnitude
greater than the weak interaction rates.
This implies that both the non-leptonic reactions have the same chemical
potential imbalance:
\begin{equation}
\delta\mu \equiv \delta\mu_n - 
\delta\mu_{\Lambda} = 2\delta\mu_n - \delta\mu_p - \delta\mu_\Sigma. 
\end{equation}
Assuming small perturbations and using the neutron fraction $x_n$, this
yields the relation
\begin{eqnarray}
{\delta \mu\over n_B\delta x_n} &=& \alpha_{nn} +
{(\beta_n-\beta_\Lambda)(\alpha_{np}-\alpha_{\Lambda p}+\alpha_{n\Sigma}
-\alpha_{\Lambda\Sigma})\over
2\beta_\Lambda-\beta_p-\beta_\Sigma}\nonumber\\
&&-\alpha_{\Lambda n}
-{(2\beta_n-\beta_p-\beta_\Sigma)(\alpha_{n\Lambda}-\alpha_{\Lambda\Lambda})
\over 2\beta_\Lambda -\beta_p-\beta_\Sigma},\label{delta1}
\end{eqnarray}
where $\beta_i$ is given by
\begin{equation}
\beta_i = \alpha_{ni} + \alpha_{\Lambda i} - \alpha_{pi} -
\alpha_{\Sigma
i}.
\end{equation}
and $\alpha_i$ is given by
\begin{equation}
\alpha_{ij} = \left( \partial \mu_i \over \partial n_j \right)_{n_k,
{\scriptscriptstyle k\ne j}}.
\end{equation}
These expressions apply for densities where the $\Lambda$ and $\Sigma^-$
hyperons are both present.
In the regime where only the $\Sigma^-$ hyperon is present the strong
interaction constraint is no longer a factor and one obtains a simpler
form
\begin{eqnarray}
{2\delta \mu\over n_B\delta x_n} &=&
4\alpha_{nn}-2(\alpha_{pn}+\alpha_{\Sigma n}+\alpha_{np}+\alpha_{n\Sigma})
\nonumber\\
&&+\alpha_{pp}+\alpha_{\Sigma p}
+\alpha_{p\Sigma}+\alpha_{\Sigma\Sigma}.\label{delta2}
\end{eqnarray}
LO used numerical differencing to calculate the derivatives $\alpha_{ij}$,
but due to a coding error the $\sigma$ meson field was not differenced
properly.
The maximum error this induces in the value of $\delta\mu/{n_B \delta x_n}$
is less than $0.3\%$, well below the uncertainties of the problem.

\subsection{Superfluidity}

In the temperature range of interest (below $10^{10}$~K), nucleons and
hyperons are expected to form Cooper pairs near their Fermi surfaces and
act as superfluids.
This greatly slows reaction rates and has an important effect on transport
coefficients such as bulk viscosity.

The energy associated with the Cooper pairing is given by the bandgap.
Assuming $^1S_0$ pairing, the zero-temperature bandgap $\Delta_0$ is
related to the superfluid critical temperature $T_C$ by~\cite{Ashcroft}
\begin{equation}
\label{TC}
kT_C=0.57\Delta_0.
\end{equation}
We use the LO fit to the finite-temperature bandgap
\begin{equation}
\Delta(T)=\Delta_0\left[1-\left(\frac{T}{T_C}\right)^{3.4}\right]^{0.53}
\label{gapT}.
\end{equation}

For the $\Lambda$ hyperons we use the empirical fit made by LO to the
zero-temperature gap function $\Delta_{0\Lambda}$ of the $\Lambda$ hyperon
as computed by Balberg and Barnea~\cite{Balberg:1997hs}.
The calculation by Balberg and Barnea~\cite{Balberg:1997hs} is constrained
by experiments on double $\Lambda$ hypernuclei, though like other
calculations of $\Delta_\Lambda$ it is likely only good to within a
factor of 2 or 3.
The zero-temperature gap depends on the total baryon
number density $n$ and on the Fermi momentum $\pL$ in a way that LO found
was well fit by
\begin{eqnarray}
\Delta_{0\Lambda} \left(\pL,n\right)&=&5.1{\pL}^3 
\left(1.52-\pL\right)^3\nonumber\\
&&\times\left[0.77+0.043\left(6.2n-0.88\right)^2\right].
\label{gapLambda}
\end{eqnarray}
Here $\Delta_{0\Lambda}$ is in MeV, $\pL$ in fm$^{-1}$, and $n$ in
fm$^{-3}$.
Thus the critical temperature for $\Lambda$ superfluidity peaks somewhat
below $10^{10}$~K.

The $\Sigma^-$ superfluid bandgaps are not as well known as
$\Delta_\Lambda$ due to the absence of similar experiments on hypernuclei
containing the $\Sigma^-$ hyperon.
Determination of the hypernuclei energy levels for the $\Sigma^-$ hyperon
are constrained by the very short decay time of the hyperon.
Takatsuka et al~\cite{Takatsuka:2001tt} have however calculated, using
several models of the nuclear interaction, that the bandgap lies in the
range $\Delta_\Lambda\leq\Delta_{{\Sigma}^-}\leq10\Delta_\Lambda$.
To account for the uncertainty in the bandgap we perform our
calculation for two cases as in LO: when
$\Delta_{\Sigma^-}=\Delta_{\Lambda}$ and when
$\Delta_{\Sigma^-}=10\Delta_{\Lambda}$.

For the superfluidity of protons we use model 2p of Ref.~\cite{Kaminker},
which is one of the bandgap models used by HLY.
(LO did not model proton superfluidity, since it was not relevant at the
temperatures greater than $10^{10}$~K that they considered.)
This bandgap model is not based on a specific calculation of the bandgap
but preserves the general features of bandgaps predicted by various
microscopic theories.
In this model the zero temperature gap depends only on the Fermi momentum
$k_p$ of the proton.
The relation for $T_C$ is given by
\begin{equation}
T_C=T_0\left[\frac{k_p^2}{k_p^2+k_1^2}\right]
\left[\frac{(k_p-k_2)^2}{(k_p-k_2)^2+k_3^2}\right],
\end{equation}
where $k_1=1.117$ fm$^{-1}$, $k_2=1.329$~fm$^{-1}$, $k_3=0.1179$~fm$^{-1}$
and $T_0=17\times10^9K$.
The Fermi momentum $k_p$ is in fm$^{-1}$ units and $T_C$ is non-zero only
if $k_p<k_1$.
This leads to $T_C$ for the protons peaking somewhat below $10^{10}$~K,
similar to the $\Lambda$ critical temperature.

Since neutrons are, after all, more abundant than other particles in
neutron stars, they have higher energies and at supernuclear densities are
unlikely to form $^1S_0$ pairs.
They can have a $^3P_2$ pairing as considered in Ref.~\cite{Kaminker}, but
it is much weaker and corresponds to a maximum critical temperature
$T_C\simeq3 \times10^8$~K.
Below this temperature the contribution to the critical frequency curve
from hyperon bulk viscosity is quite low, most of the contribution coming
from viscosity at the crust-core interface.
Therefore neutron superfluidity does not much affect the viability of
persistent gravitational wave emission from the $r$-modes, and we neglect
it here.

The effect of superfluidity on the microscopic relaxation
timescale~(\ref{tau1}) or~(\ref{tau2}) appears, after doing the lengthy
collision integrals (see HLY and LO), as multiplicative factors in the
average matrix elements:
\begin{eqnarray}
\langle |\mathcal{M}_\Sigma|^2 \rangle &\to& R_{nnp\Sigma} \langle
|\mathcal{M}_\Sigma|^2 \rangle,
\\
\langle |\mathcal{M}_\Lambda|^2 \rangle &\to& R_{npp\Lambda} \langle
|\mathcal{M}_\Lambda|^2 \rangle.
\end{eqnarray}
If none of the participating particle species is superfluid, an $R$ factor
is 1.
Each $R$ factor is reduced as more species become superfluid.
Our treatment of these factors is based on that used by HLY, who calculate
$R$ for the cases when either one or two of the particles participating in
the reaction are superfluid.
HLY do not however provide a $3$-particle $R$ which is required for
calculating the timescale of the $\Lambda$ reaction~(\ref{Lamb}).
Our numerical investigations found 
(\ref{Crit_freq_curves}) that expressing 
$R_{nnp\Sigma}$ as a
product of two single-particle factors $R_pR_{\Sigma^-}$ instead of HLY's
two-particle factor does not affect the $r$-mode critical frequency much.
Therefore throughout this paper we express all $R$'s as products of the
appropriate single-particle factors.
The expression for the single-particle $R$ from Eq.~(30) of HLY is
\begin{equation}
R=\frac{a^{5/4}+b^{1/2}}{2}\exp \left(0.5068-\sqrt{0.5068^2+y^2} \right),
\end{equation}
where $y=\Delta(T)/kT$, $a=1+0.3118y^2$ and $b=1+2.566y^2$. 
LO used a simple factor $e^{-y}$ for the hyperons only, since they were
focusing on the ordinary fluid case.
The exponential is indeed the dominant behavior of the HLY expression for
$y\gg1$, but the HLY expression is significantly more accurate for $y$ less
than about 5--10, which turns out to be the astrophysically interesting
region.

\begin{figure}
\includegraphics[width=2.8in]{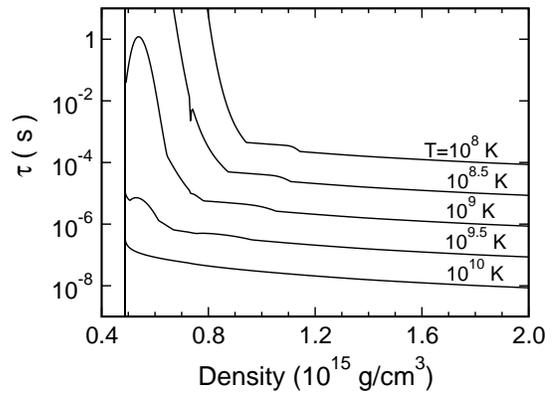}
\caption{
\label{tauofrho}
Microscopic relaxation timescale as a function of density for various
temperatures.
The equation of state is H3 and the $\Sigma^-$ and $\Lambda$ bandgaps are
assumed equal.
Thus at the highest temperature plotted, no superfluid effects are present.
}
\end{figure}

\begin{figure}
\includegraphics[width=2.8in]{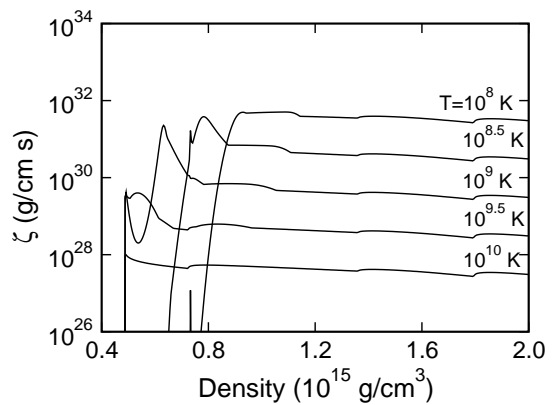}
\caption{
\label{zetaofrho}
Bulk viscosity coefficient as a function of density for various
temperatures, for the same model as Fig.~\ref{tauofrho}.
The star is rotating with angular frequency 2300 radians per second.
}
\end{figure}

We plot the most important results of our microphysics calculations.
Figure~\ref{tauofrho} shows a typical relaxation timescale as a function of
density, and Fig.~\ref{zetaofrho} shows the corresponding bulk viscosity
coefficient.
Among the equations of state we use, H3 has a relaxation timescale in the
middle of the distribution.
Neglecting superfluidity, the timescales are about a factor of 40 shorter
than in LO.
About 20 of this comes from making the matrix elements consistent with the
quasiparticle picture, and most of the rest comes from changes to the
equation of state (including accounting for the typos in
Ref.~\cite{Glendenning:1985jr} which propagated to LO).
Fig.~\ref{zetaofrho} is reduced from LO by a further factor of 2 due to
the relativistic factor $\gamma_\infty - \gamma_0$.
Including superfluidity, the shapes of the curves are different from LO but
the peak timescales are comparable.
It is difficult to compare to HLY because at these temperatures the
high-frequency approximation they use ($\hat\omega \tau \gg 1$) does not
hold.

\section{Macrophysics}

With the bulk viscosity coefficient $\zeta$ in hand, we can proceed to
compute $r$-mode damping times and critical frequencies.
For this we also need the hydrodynamic structure of the $r$-mode as well as
equilibrium models of the structure of a neutron star (i.e. its
density profile).

\subsection{Stellar structure}

We are motivated by the sensitivity of the hyperon population to the
central density to include the effects of general relativity and rotation
in our stellar models.
Relativistic stellar models have higher densities than nonrelativistic
ones, and rotating models have lower densities than nonrotating ones.
LO used relativistic, nonrotating stellar models; while other work such as
HLY and Ref.~\cite{vanDalen:2003uy} did not treat the stellar or mode
structure in much detail.
Since we are interested in exploring a broader parameter space, we do not
restrict ourselves to nonrotating stars.

We take into account the effect of rotation on the neutron star structure
using Hartle's slow-rotation approximation~\cite{Hartle:1967he}.
Hartle's formalism is based on treating a rotating star as a perturbation 
on a non-rotating star.
We start by solving the Oppenheimer-Volkoff equations for a static
spherically symmetric star in a form due to Lindblom~\cite{Lindblom:1992}
\begin{eqnarray}
\frac{dm}{dh} &=& -\frac{ 4\pi\epsilon(h)r(h)^3 [r(h)-2m(h)]} {m(h) + 4\pi
r(h)^3p(h)},
\\
\frac{dr}{dh} &=& -\frac{r(h) [r(h)-2m(h)]} {m(h) + 4\pi r(h)^3p(h)},
\end{eqnarray}
using as independent variable the specific enthalpy
\begin{equation}
h(p) = \int_0^p dp' / [p' + \epsilon(p')].
\end{equation}
From $r(h)$, the radius at which the enthalpy is $h$, and $m(h)$, the mass
contained within a sphere of that radius, we find $m(r)$.
We also find $\nu(r)$, the logarithm of the time-time metric component
which is based on $h(r)$.
We then use this solution of the static star to solve for a rotating star
with the same central pressure.

The metric of a slowly rotating star to second order in $\Omega$, the spin
frequency, can be written as
\begin{eqnarray}
\nonumber ds^2 
= 
&-&e^\nu \left[1 + 2(h_0+h_2P_2) \right] dt^2\\
&+& 
\nonumber \left[1 + \frac{2(m_0+m_2P_2)}{(r-2M(r))} \right] \left[1- 
\frac{2M(r)}{r} \right]^{-1}
dr^2\\
\nonumber &+& 
r^2\left[1+2(v_2-h_2)P_2\right] \left[d\theta^2+\sin^2 
\theta (d\phi-\omega dt)^2\right]\\
&+& O(\Omega^3).
\label{rot_metric}
\end{eqnarray}
Here $P_2=P_2(\cos\theta)$ is the 2nd order Lagrangian polynomial.
The quantity $\omega$ is the frame dragging frequency that is
proportional to $\Omega$, while $h_0$, $h_2$, $m_0$, $m_2$, and $\nu_2$
are all functions of $r$ that are proportional to $\Omega^2$.
Before we solve the Einstein equations for all the subscripted quantities
in Eq.~(\ref{rot_metric}) we need to solve for $\bar \omega = 
\Omega-\omega$.
This is done by solving
\begin{equation}
\frac{1}{r^4} \frac{d}{dr} \left(r^4 j(r) \frac{d \bar \omega(r)}{dr} 
\right) +\frac{4}{r} \frac{d j(r)}{d r} \bar\omega(r)=0,
\label{framedrag}
\end{equation}
with the definition
\begin{equation}
j(r)=\exp^{-\nu/2}\left[1-\frac{2 M(r)}{r}\right]^{1/2},
\label{jay}
\end{equation}
and the boundary conditions that at the center of the star
$\bar\omega(0)=\bar\omega_i$, an arbitrary constant, and
$(d\bar\omega(r)/dr)_{r=0}=0$.
From Eq.~(\ref{framedrag}) one determines $\Omega$ and the angular
momentum $J$ corresponding to $\bar\omega_i$ as follows:
\begin{eqnarray}
\label{ang_mom}
J&=& \frac{1}{6}R^4 \left(\frac{d\bar\omega(r)}{dr}\right)_{r=R},\\
\Omega&=&\bar\omega (R)+ \frac{2J}{R^3}.
\end{eqnarray}
Here $R$ is the radius of the star.
To obtain the desired value of $\Omega$ one can scale $\bar\omega$ thus:
\begin{equation}
\bar\omega(r)_{new} = \bar\omega(r)_{old}(\Omega_{new}/\Omega_{old}). 
\end{equation}
The angular momentum $J$ corresponding to $\Omega_{new}$ can then be
calculated using Eq.~(\ref{ang_mom}).

A measure of the change in the star structure caused by rotation is
$\xi(r,\theta)$ which represents the displacement of constant energy
density surfaces between the rotating star and the corresponding 
non-rotating star.
This displacement's angular dependence can be separated out:
\begin{equation}
\xi(r,\theta)=\xi_0(r)+\xi_2(r)P_2
\label{xi_eq}.
\end{equation}
The $l=0$ part of this deformation, $\xi_0$, can be obtained by solving
the following equations:
\begin{eqnarray}
\label{rot_eq_1}
\nonumber \frac{dm_0(r)}{dr} = &+&4\pi r^2 \frac{d\epsilon(r)}{dp(r)} 
\left[\epsilon(r)+p(r)\right]p_0^*(r)\\
\nonumber 
&+&\frac{1}{12}j^2(r)r^4\left(\frac{\bar\omega(r)}{dr}\right)^2\\
&-&\frac{1}{3}r^3 \frac{dj^2(r)}{dr} \bar\omega^2(r),\\
\label{rot_eq_2}
\nonumber \frac{dp_0^*(r)}{dr} = &-&\frac{m_0\left[1+8\pi r^2 
p(r)\right]}{r-2M(r)}^2\\
\nonumber 
&-&\frac{4\pi 
\left[\epsilon(r)+p(r)\right]r^2}{r-2M(r)}p_0^*(r)\\
\nonumber &+& \frac{r^4j^2(r)}{12\left[r-2M(r)\right]}\left( 
\frac{d\bar\omega(r)}{dr}\right)^2\\
 &+&\frac{1}{3} \frac{d}{dr} \left(\frac{r^3j^2(r)\bar\omega^2(r)} 
{r-2M(r)} 
\right),
\end{eqnarray}
where $p_0^*$ is a pressure perturbation, the density profile $\epsilon(r)$
is the energy density of the non-rotating star as function of $r$, and
$p(r)$ is the pressure of the star as a function of $r$.  
The boundary conditions for the above equations are that both $m_0$ and
$p_0^*$ vanish at the origin.

The gravitational mass of the rotating star is given by 
\begin{equation}
M'(R)=M(R)+m_0(R)+\frac{J^2}{R^3},
\label{mass_rot}
\end{equation}
which is greater than the mass of the nonrotating star with the same
central pressure.
We wish to build sequences of stars rotating at various rates all with the
same gravitational mass.
To do this we first solve the Oppenheimer-Volkoff equations; then, solve
Eq.~(\ref{framedrag}), (\ref{rot_eq_1}), and (\ref{rot_eq_2}); then
calculate the new mass of the rotating star using Eq.~(\ref{mass_rot}).
This procedure is repeated with lower values of the central pressure until
the same mass as the nonrotating star is obtained.

To perform the viscosity calculations we need information on the structure
of the rotating star contained in its density profile.
Once the correct mass is obtained the density profile $\epsilon(r)$ is 
calculated by using
\begin{equation}
\epsilon_{rot}(r)=\epsilon_{stat}(r)-\frac{d\epsilon_{stat}(r)}{dr}\xi_0(r),
\end{equation}
where
\begin{equation}
\xi_0(r)=-p_0^*(r)\frac{\epsilon(r)+p(r)}{dp(r)/dr}.
\end{equation}
Here, $\epsilon_{rot}(r)$ is the density profile of the rotating star,
while $\epsilon_{stat}(r)$ is the density profile of the non-rotating one.
This yields only the spherically symmetric part of the density
perturbation, which is the dominant one for our purposes.

\subsection{Driving and damping timescales}

The stability of an $r$-mode is determined by calculating the damping and
driving timescales.
The mode is unstable if the driving timescale is shorter than the viscous
damping timescale.
This can be expressed in terms of an overall $r$-mode timescale $\tau_r$
such that
\begin{equation}
\frac{1}{\tau_r(\Omega,T)}=\frac{1}{\tau_{GR}(\Omega)}+\frac{1}{\tau_V(\Omega,T)},
\label{net_timescale}
\end{equation}
where $\tau_{GR}<0$ is the damping timescale of the mode due to
gravitational radiation and $\tau_V>0$ is the damping timescale resulting
from the sum of all viscous processes in the star.
The star's spin angular frequency and temperature are represented by
$\Omega$ and $T$, respectively.
A mode unstable to gravitational radiation corresponds to $\tau_{r}<0$ in
Eq.~(\ref{net_timescale}), and the critical frequency $\Omega_c$ of the
star as a function of temperature is found by solving
$1/\tau_r(\Omega_c,T)=0$.

The gravitational radiation timescale is given by
\begin{equation}
{1\over\tau_{GR}} = -{1\over2\tilde{E}} \left( d\tilde{E} \over dt
\right)_{GR},
\end{equation}
where $\tilde{E}$ is the mode energy in a frame rotating with the star and
$(d\tilde{E}/dt)_{GR}$ is the rate at which energy is emitted in
gravitational waves in the same frame.
For the current quadrupole $r$-mode, there exist simple expressions (to
lowest order in mode amplitude and angular velocity of the star) for the
energy and gravitational radiation timescale:~\cite{Lindblom:1998wf}
\begin{eqnarray}
\tilde E &=& \frac{1}{2}\alpha^2\Omega^2 R^{-2}\int_0^R \epsilon(r) r^6 dr,
\nonumber
\\
\frac{1}{\tau_{GR}}&=&-\frac{131\,072\pi} {164\,025} \int_0^R {\epsilon(r)}
r^6\,dr.
\label{tauGR}
\end{eqnarray}
Here $\alpha$ is a dimensionless amplitude coefficient which cancels out of
the driving and damping timescales.

The viscous damping time scale $\tau_{V}$ is given by
\begin{equation}
\frac{1}{\tau_{V}}=
-\frac{1}{2\tilde{E}}\left(\frac{d\tilde{E}}{dt}\right)_V,
\label{damping_timescale}
\end{equation}
where $(d\tilde{E}/dt)_V$ is the rate at which energy is being drained from
the mode by viscosity.
Here we consider only hyperon bulk viscosity from the nonleptonic processes
discussed in the previous section.
Leptonic (Urca) processes also contribute, but in LO they were shown to be
overwhelmed by the viscosity due to nonleptonic processes.
Although the crust-core viscosity is important at low temperatures---it 
determines the shape of the negatively sloped part of the instability 
curve which in turn puts a constraint on how much farther along 
the temperature axis the rising part of the instability curve should be 
located at to prevent a thermal runaway---we neglect it.
For a reasonable estimate of the shape of the crust-core viscosity curve
and of the value of the saturation amplitude the stars temperatures hould
increase to around $10^9$~K~\cite{Heyl:2002pe}.
To explore the effect of the crust-core viscosity and the saturation 
amplitude in more details is beyond the scope of this paper. 
We can then compute the viscous damping timescale as in LO:
To lowest order in $\Omega$, we can write
\begin{equation}
\left(\frac{d\tilde{E}}{dt}\right)_V=-4\pi\int_0^R \zeta(\epsilon(r))
\langle|\vec{\nabla} \cdot \delta\vec{v}|^2\rangle r^2\,dr,
\label{dEbydt}
\end{equation}
where $\langle|\vec{\nabla} \cdot \delta\vec{v}|^2\rangle$ is the angle
average of the square of the hydrodynamic expansion. 
LO found that the expansion found numerically in
Ref.~\cite{Lindblom:1999yk} could be fit well by
\begin{equation}
\langle|\vec{\nabla} \cdot \delta\vec{v}|^2\rangle = \frac{\alpha^2 
{\Omega}^2} 
{690}
\left(\frac{r}{R}\right)^6\left[1+0.86\left(\frac{r}{R}\right)^2\right]
\left( \Omega^2 \over \pi G\bar\epsilon \right)^2,
\label{expansion}
\end{equation}
where $\bar{\epsilon}$ is the mean density of the (nonrotating) star.
[In Eq.~(6.6) of LO the last factor is missing due to a misprint, but is
properly included in the calculations.]
This approximation breaks down for $r/R\approx1$, but since the bulk
viscosity is coming from the hyperon core that is enough.
In Table~\ref{radii} we give typical radii of nonrotating stars and their
hyperon cores.

\begin{table}
\begin{ruledtabular}
\begin{tabular}{lcccc}
EOS & $\epsilon_c$ ($10^{14}$g/cm$^3$) & $R$ (km) & $R_\Sigma$ (km) & 
$R_\Lambda$ (km)
\\
\hline
H3	& 6.86	& 13.74	& 6.38	& 0.00
\\
H4 	& 6.12	& 13.85	& 4.50	& 0.00
\\
H5 	& 7.70	& 13.33	& 5.37	& 0.00
\\
H6 	& 7.84	& 13.44	& 6.70  & 2.46
\\
H7	& 9.00	& 13.01  & 6.03  & 0.00
\end{tabular}
\end{ruledtabular}
\caption{
\label{radii}
Parameters of non-rotating $1.4M_\odot$ equilibrium stars for each equation
of state (EOS).
The central mass density is $\epsilon_c$, $R$ is the stellar radius,
$R_\Sigma$ is the distance out to which $\Sigma^-$ hyperons appear, and
$R_\Lambda$ is the corresponding distance for $\Lambda$ hyperons.
}
\end{table}

\subsection{Critical frequency curves}
\label{Crit_freq_curves}

We are now in a position to solve for the critical frequency as a function
of temperature for various neutron star models.
This must be done numerically.

Because of the inclusion of rotation the process is somewhat more involved 
than
in LO.
The critical frequency curve is calculated in two stages to ensure that the
correct gravitational mass is retained after including the corrections to
it from rotation.
In the first stage we calculate the critical frequency as a function of
temperature for a nonrotating star using the Oppenheimer-Volkoff formalism.
Then a set of points on this curve is taken.
In the second stage, for each point we find the structure (density profile)
of a star rotating at that frequency with the same gravitational mass as
the nonrotating star.
Since Hartle's formalism produces stars with greater gravitational masses,
this step requires several iterative trials of the central pressure to find
the star with the same gravitational mass.
Using the new density profile with rotation, we keep the frequency fixed
and use the corresponding temperature on the nonrotating curve as an
initial guess for a routine that finds the root of $1/\tau_r$, thereby
obtaining the correct temperature for the marginally star rotating at that
frequency.
Because of this it is cumbersome to trace out parts of the curves where the
critical frequency decreases with temperature, and since those parts of the
bulk viscosity curves are not important for determining the viability of
persistent gravitational wave emission we neglect them.

\begin{figure}
\includegraphics[width=2.8in]{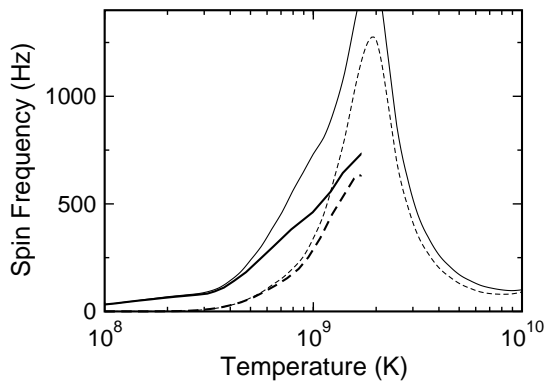}
\caption{
\label{f:rotation_effect}
Critical frequency curves 1.4M$_\odot$ neutron stars with equal bandgaps
for $\Sigma^-$ and $\Lambda$ hyperons.
The dashed curve is for equation of state H3 and the solid curve is for H6.
The thinner curves neglect the effect of rotation on stellar structure
while the thicker curves include it.
The curves including rotation are extended to their peaks, but no further.
}
\end{figure}

The importance of rotation is shown in the critical frequency curves
plotted in Fig.~\ref{f:rotation_effect}.
(Kepler frequencies for these equations of state are 750--850~Hz for
1.4~$M_\odot$ stars.)
Equation of state H6 is soft and has the highest hyperon population for a
given mass, and thus its curve is higher than for H3 which is intermediate.
This plot shows the general trend that the effect of rotation is more
significant for softer equations of state.
For persistent gravitational wave emission the key question is whether or
not a neutron star undergoing thermal runaway will be blocked by hitting
the curve at a temperature lower than a few times $10^9$~K.
Even for the soft H6 equation of state, the effect of rotation on stellar
structure changes the temperature at which a star hits the curve by no more
than a factor of 2 (still within the region that is good for gravitational
wave emission).
More significant is the fact that the frequency at which the curve peaks
can be considerably reduced from the value neglecting rotation.
If the peak is low enough, thermal runaway can still occur---for example in
this plot it can occur for H3 stars if they hit above 700~Hz---and the
outlook is bad for persistent gravitational wave emission.

\begin{figure}
\includegraphics[width=2.8in]{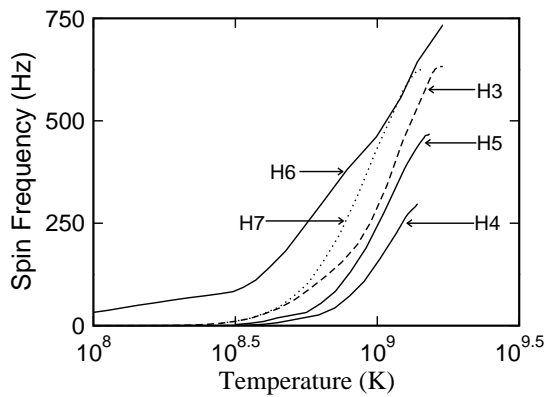}
\caption{
\label{f:M14prop}
Critical frequency as a function of temperature for 1.4~$M_\odot$ stars
with different equations of state.
All curves assume the superfluid bandgaps of the $\Lambda$ and $\Sigma^-$
hyperons are equal.
}
\end{figure}

In Fig.~\ref{f:M14prop} we show the effect of the equation of state on the
critical frequency curve for 1.4~$M_{\odot}$ stars.
The curves generically reach their peak at temperatures of
1--2$\times10^9$~K, which is good for persistent gravitational wave
emission.
However, the heights (frequencies) of the peaks change significantly.
The peak is highest for H6, which has the largest hyperon population.
It is almost identical for H3 and H7, although those two equations of state
are the furthest separated in the $(K,m^*,x_\sigma)$ parameter space.
This seems to be due to the similar radii of their hyperon-containing cores
as in Table~\ref{radii}.
All three of these equations of state then allow persistent gravitational
wave emission in 1.4~$M_\odot$ stars.
The peak is lowest for H4, which is not surprising since that is the
stiffest equation of state and has the lowest hyperon population.
For H4 the peak is so low that, depending on the details of the
crust-related damping, the full curve including that damping might show no
significant effect from hyperon bulk viscosity---it might decrease
throughout the temperature range and never allow persistent emission.
In any case, for H4 and H5 persistent emission is not possible for the most
rapidly rotating stars in LMXBs (up to 619~Hz) if they are 1.4~$M_\odot$.
This demonstrates in principle how observation of $r$-mode gravitational
waves from an LMXB could rule out some equations of state or, more broadly,
constrain hyperon populations.

\begin{figure}
\includegraphics[width=2.8in]{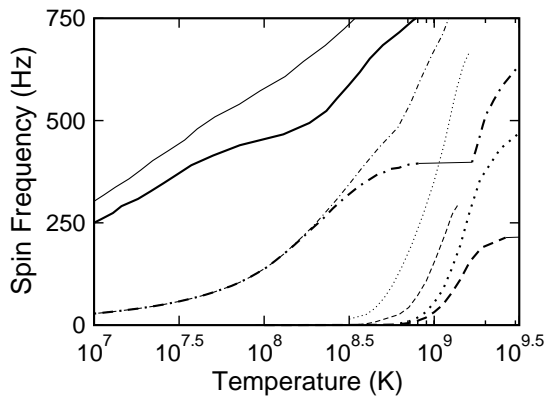}
\caption{
\label{f:K300m70x72prop}
Instability curves for H4 stars with different masses.
The thicker curves use the larger $\Sigma^-$ bandgap while the 
thinner curves use the smaller one.
The solid lines represent $2.0M_\odot$ stars, the dot-dash lines repsents
$1.8M_\odot$ stars, the dotted line represents $1.6M_\odot$ stars, and the
dashed line represents $1.4M_\odot$ stars.
}
\end{figure}

In Fig.~\ref{f:K300m70x72prop} we plot critical frequency curves for stars
with different masses for the H4 equation of state, which is the stiffest
of those we use.
(For other equations of state, the stellar mass affects the curve more but
is constrained to a narrower range, so the overall variation is greatest
for H4.)
We also vary the $\Sigma^-$ bandgap from the $\Lambda$ value to 10 times
that value.
It can be seen that increasing the mass pushes the curve to lower
temperatures and raises its peak.
This is due to the fact that higher mass stars have higher central
densities and hence, in absolute terms, larger hyperon populations than
their lower mass counterparts.
Persistent $r$-mode gravitational wave emission is viable in all but the
1.4~$M_\odot$ neutron stars.
The temperature at which this can happen is strongly dependent on the mass,
and to a lesser but still significant extent (factor 3) on the $\Sigma^-$
bandgap, with higher bandgaps leading to higher temperatures at a given
frequency.
(The small horizontal sections of the $1.4M_\odot$ and $1.6M_\odot$ large
bandgap curves should actually dip slightly.
This behavior is missed due to our method of including rotation, but is
unimportant for the thermal runaway and persistent gravitational wave
emission.)

The curves for 1.8--2.0~$M_\odot$ stars are particularly interesting in
light of predicted interior temperatures of neutron stars in LMXBs.
These can be determined from observations in quiescence of the surface
temperatures of three slowly accreting neutron stars mentioned in Brown,
Bildsten, and Chang~\cite{Brown:2002rf}.
To calculate the interior temperatures we use Fig.~2 from Yakovlev et
al~\cite{Yakovlev:2003ed}, which depicts the relationship between the core
and the measured surface temperatures using different atmospheric and
equation of state models.
We thereby obtain a temperature range ($1.5\times10^7$--$4\times10^8$K)
which covers LMXBs whose thermal emission spectra have been observed.
This result is probably statistically biased, since LMXBs with higher
accretion rates are likely to have higher temperatures but surface
temperatures are harder to observe because of the accretion discs.
At any rate, the most massive H4 stars have critical frequency curves which
would allow gravitational wave emission scenarios~\cite{Wagoner:2002vr,
Reisenegger:2003cq} to operate in that temperature range.
Less massive stars allow it at higher temperatures, which may be consistent
with the more rapidly accreting neutron stars such as Sco X-1 whose
temperatures are poorly known.

\begin{figure}
\includegraphics[width=2.8in]{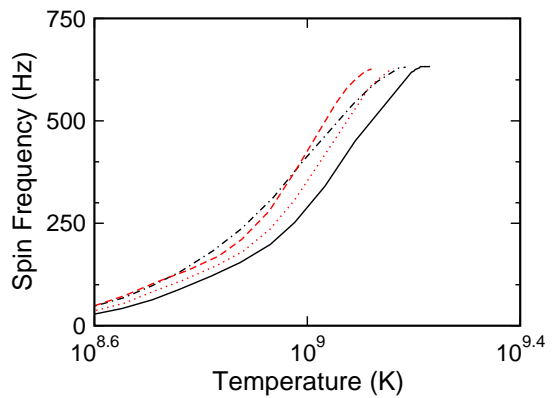}
\caption{
\label{f:300K70M14_CritFreq_var}
Variations on the critical frequency curve for a 1.4~$M_\odot$ H3 star with
equal $\Lambda$ and $\Sigma^-$ bandgaps.
The solid curve is our standard model.
The dotted curve uses asymptotic values for the axial couplings instead of
vacuum at-rest values.
The dot-dash curve uses the 2-particle superfluid reduction factor for the
$\Sigma^-$ interaction instead of two 1-particle factors.
The dashed curve simulates the fast timescales predicted by van Dalen and
Dieperink.
}
\end{figure}

The effects of varying some additional microphysical parameters are shown
in Fig.~\ref{f:300K70M14_CritFreq_var}, which depicts curves for
1.4~$M_\odot$ neutron stars with the H3 equation of state and the
$\Sigma^-$ superfluidity bandgap equal to that of the $\Lambda$.
The variations include using the asymptotic (quark) values for the
axial-vector couplings in the reaction rates, using HLY's two-particle
superfluid reduction factor instead of the product of two one-particle
factors, and using relaxation times an order of magnitude faster from van
Dalen and Dieperink~\cite{vanDalen:2003uy}.
All of these curves peak at almost identical frequencies, and the
temperature at a fixed frequency changes by less than a factor of 1.5
Therefore, the changes to these microphysical parametrs do not affect our 
earlier conclusions. 
From an astrophysical point of view, the most sensitive unknowns in this
complicated calculation are the hyperon population (thus the equation of
state and stellar mass) and superfluid bandgaps.
At a very simple level this makes sense because the hyperons appear only
after a threshold density and the bandgaps appear in exponential factors
rather than power laws.

\section{Discussion}

We have extended previous investigations of $r$-modes in accreting hyperon
stars in LMXBs as persistent sources of gravitational waves, focusing on
the bulk viscosity which is needed to prevent thermal runaway.
We have used improved microphysics compared to previous treatments, and
have accounted for the most important macroscopic correction due to
rotation of the star.

We find that persistent gravitational wave emission is quite robust.
Even the stiffest hyperonic equations of state in relativistic mean field
theory produce enough damping to stop the runaway and persistently radiate,
although some of them require neutron stars somewhat more massive than
1.4~$M_\odot$.
Stars below about 1.3~$M_\odot$ are not likely to be persistent sources
regardless of the equation of state.
The mass thresholds are somewhat more favorable for lower superfluid
bandgaps than for higher bandgaps.
Other details of the microphysics are found to be considerably less
important.
Our results seem robust for typical values of the crust-core viscosity 
and of saturation amplitude of $r$-modes, though this issue requires 
further investigation.
Stars with high masses and stiff equations of state could exist in thermal
and torque equilibrium at temperatures down to $10^7$~K.

One possible avenue for substantial improvement is the hydrodynamics.
We used a fluid expansion in the dissipation integrals which is a
reasonable approximation for a Newtonian normal fluid, but not for a
superfluid which can be much more complicated due to multiple components
and entrainment.
Work is underway to deal with the superfluid problem in general
(\cite{Andersson:2005pf} and references therein).

\begin{acknowledgments}

This work was supported by the National Science Foundation under grants
PHY-0245649 and PHY-0114375 (the Penn State Center for Gravitational Wave
Physics). We are especially grateful to Lee Lindblom for many helpful
discussions, including confirming errors in previous work.  We also thank
Lars Bildsten, Ian Jones and Dimitri Yakovlev for helpful discussions.

\end{acknowledgments}

\bibliography{paper}

\end{document}